\documentclass{iopart}
\usepackage{iopams} 
\usepackage{mathptmx}
\usepackage[utf8]{inputenc} 
\usepackage{graphicx}
\usepackage[numbers,comma,square,sort&compress,merge]{natbib} 
\usepackage{xspace}
\usepackage[
,textwidth=17.5cm
,textheight=23.5cm
,verbose
,dvips
]{geometry}
\usepackage{url}

\begin{document}
\title[Localization and defects in (In,Ga)N/GaN NWs investigated by spatially-resolved luminescence spectroscopy]{Localization and defects in axial (In,Ga)N/GaN nanowire heterostructures investigated by spatially-resolved luminescence spectroscopy}
\author{Jonas L\"ahnemann, Christian Hauswald, Martin W\"olz, Uwe Jahn, Michael Hanke, Lutz Geelhaar, Oliver Brandt}
\ead{laehnemann@pdi-berlin.de}
\address{Paul-Drude-Institut für Festkörperelektronik,
Hausvogteiplatz 5--7, D-10117 Berlin, Germany}


\begin{abstract}
(In,Ga)N insertions embedded in self-assembled GaN nanowires are of current interest for applications in solid-state light emitters. Such structures exhibit a notoriously broad emission band. We use cathodoluminescence spectral imaging in a scanning electron microscope and micro-photoluminescence spectroscopy on single nanowires to learn more about the mechanisms underlying this emission. We observe a shift of the emission energy along the stack of six insertions within single nanowires that may be explained by compositional pulling. Our results also corroborate reports that the localization of carriers at potential fluctuations within the insertions plays a crucial role for the luminescence of these nanowire based emitters. Furthermore, we resolve contributions from both structural and point defects in our measurements.
\end{abstract}

\pacs{78.67.Uh,
78.60.Hk,
81.07.St
}

\submitto{\JPD}

\twocolumn

\maketitle

\section{Introduction}
Quantum well (QW) structures of the ternary semiconductor (In,Ga)N are at the heart of solid-state lighting applications \cite{Pimputkar_natphot_2009}.
Since recent years, a number of groups are investigating (In,Ga)N insertions embedded in arrays of self-assembled, bottom-up grown GaN nanowires (NWs) \cite{Kikuchi_jjap_2004,Kishino_procspie_2007,Bavencove_nt_2011,Armitage_nt_2010, Nguyen_nl_2011,Lin_apl_2010,Chang_apl_2010,Wolz_nt_2012,Guo_apl_2011, Kehagias_nt_2013}. The NW geometry is associated with an improved crystal quality, a better elastic strain relaxation at the heterointerfaces and an enhanced light extraction efficiency compared with planar layers. Several of these groups have presented a certain degree of color tunability of such NW light emitters \cite{Chang_apl_2010,Lin_apl_2010,Wolz_nt_2012,Guo_apl_2011,Sekiguchi_apl_2010}. Nevertheless, in all these studies, the (In,Ga)N emission band is broader than for planar QWs. The latter has led several groups to propose devices with the potential for direct (phosphor-free) emission of white light \cite{Armitage_nt_2010,Lin_apl_2010,Nguyen_nl_2011,Albert_apl_2012,Guo_apl_2011}. However, a ``white to the eye'' emission is not sufficient for most applications. Both the color temperature and color rendering are critical for many applications, such as the motion picture business to name just one example~\cite{Academy_2014}. The ability to control the spectral properties of these structures in a reproducible fashion is therefore an important prerequisite for their commercial use. To attain the necessary control over the growth, a thorough understanding of the emission characteristics and the underlying physical processes is needed. To this end, micro-photoluminescence ($\mu$PL) spectroscopy on single NWs \cite{Kawakami_apl_2006,Bardoux_prb_2009,Lahnemann_prb_2011}, but also spatially resolved cathodoluminescence (CL) spectroscopy \cite{Lahnemann_prb_2011,Tourbot_nt_2012,Bruckbauer_nt_2013}, have been employed.  Following these approaches, we use both CL spectral mapping in an SEM and $\mu$PL to investigate (In,Ga)N insertions in GaN NWs produced by a bottom-up approach. The emission characteristics along the NW axis are resolved in 1D CL line scans, whereas the (In,Ga)N insertions are also investigated by 2D spectral maps. Furthermore, we extend the analysis to measurements at helium temperatures, and thereby we investigate the presence of defects, such as point defects and zinc-blende segments.

\section{Experimental}

In this article, a sample of GaN NWs containing six (In,Ga)N insertions is studied. The NWs and embedded heterostructures were grown by plasma-assisted molecular beam epitaxy (MBE) using the self-induced growth mode on Si(111) substrates \cite{Sanchez-Garcia_jcg_1998}. To facilitate the incorporation of In, the substrate temperature was reduced from the 780~$^\circ$C used to grow the GaN NW base to about 600~$^\circ$C. Alternating (In,Ga)N insertions and GaN barriers were grown. Care has to be taken to maintain the N-rich conditions at the growth front even for the lower desorption rate of Ga at the reduced temperature~\cite{Fernandez-Garrido_nl_2013}. Finally, the structure was capped with a GaN segment of about 30~nm without raising the temperature. The composition and dimensions of the (In,Ga)N insertions can be assessed from the superlattice fringes of the stack of (In,Ga)N insertions in X-ray diffraction (XRD) profiles \cite{Wolz_apl_2011,*Wolz_apl_2012,Wolz_cgd_2012,Wolz_nt_2012}, which yields $3\pm2$~nm for the thickness of the (In,Ga)N insertions and $7\pm2$~nm for the thickness of the GaN barriers. Their average In content is $x=0.26\pm0.1$. Note that these are ensemble averages and the large error margins reflect the statistical distribution of dimensions and composition between different NWs of the NW ensemble. Transmission electron microscopy (TEM) on similar samples revealed that the insertions are laterally embedded in a GaN shell and further confirmed that the barriers are indeed significantly wider than the insertions \cite{Wolz_nt_2012}. The specific sample under investigation was selected for the detailed analysis using luminescence spectroscopy by virtue of its high luminescence yield. In comparison to our previous study in \cite{Lahnemann_prb_2011}, this sample contains thinner insertions separated by thicker barriers.

Cathodoluminescence spectroscopy and imaging was performed using a Gatan Mono-CL3 system mounted to a Zeiss Ultra55 field-emission scanning electron microscope (SEM). The light emitted by the sample was collected by a parabolic mirror and then focused onto the entrance slit of a 300~mm Czerny-Turner grating spectrometer. For monochromatic imaging, the light from the exit slit of the spectrometer was focused on the active area of a photomultiplier. Alternatively, for spectral imaging, the diffracted light could be focused onto a liquid nitrogen cooled charge-coupled device array. To accommodate the wide spectral bandwith of the (In,Ga)N emission on the CCD, a grating with 300~lines/mm blazed at 500~nm was employed. The SEM was operated at an acceleration voltage ($V_\mathrm{acc}$) of 3--5~kV. For measurements on single NWs, these were dispersed on a piece of Si substrate.

For $\mu$PL measurements, single NWs were detached and dissolved in 2-propanol in an ultrasonic bath and subsequently deposited onto a TEM finder grid supported by a carbon film and placed on a Si substrate. The (In,Ga)N insertions were resonantly excited using the 413.1~nm line of a Kr$^+$ laser. The beam was focused to a spot with 1~$\mu$m diameter and the PL was collected in a confocal geometry and detected with a spectral resolution of 1~meV. Using optical density filters, the incident excitation density was varied over three orders of magnitude with a maximum excitation of $I_0=500$~kW/cm$^2$. Note that due to the comparably small volume of (In,Ga)N, the absorbed intensity can be expected to be much lower.

\begin{figure*}
\centering
\includegraphics*[width=16cm]{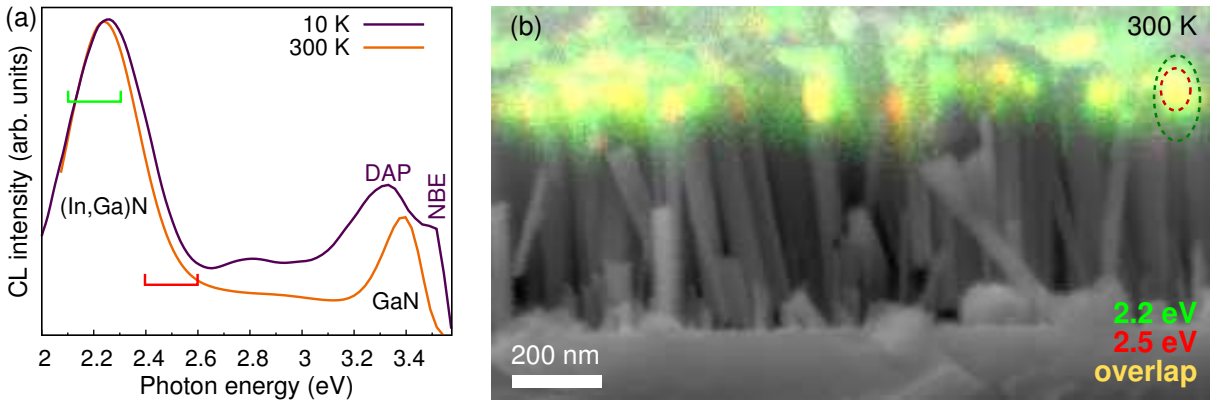}
\caption{\label{fig:ingan_m81310ens}(a) Normalized CL spectra recorded in top-view geometry at 10 and 300~K on the NW ensemble. The horizontal bars indicate the spectral windows around 2.2 (green) and 2.5~eV (red) used for the cross-sectional false-color CL images of the (In,Ga)N luminescence shown in (b) superimposed on the corresponding SEM image. A spatial overlap of emission at both energies leads to the yellow color. For the CL images, a wide spectral bandpass of 50~nm ($\approx 200$~meV) was selected to accommodate the rather wide luminescence band. Carriers diffusing to the insertions recombine at the lowest available energies, whereby the emission at higher energies originates from a spatially more confined region, as marked by the dashed lines.}
\end{figure*}

\section{Results and Discussion}

\subsection{Ensemble emission}

A comparison of the CL spectra from this NW ensemble measured at temperatures of 300 and 10~K is given in figure~\ref{fig:ingan_m81310ens}(a). The area scanned by the electron beam of 360~$\mu$m$^2$ includes around 20\,000 NWs. Both the emission from the GaN NWs and the (In,Ga)N insertions are visible. The latter is centred in the green spectral region at 2.24~eV for 300~K and at 2.25~eV for 10~K. The full widths at half maximum (FWHM) are 340~meV and 380~meV, respectively. Hereafter, this emission is referred to as the (In,Ga)N band. The rather large FWHM is comparable to that reported in various literature reports of PL or EL from other MBE grown (In,Ga)N insertions in GaN NWs, the central energies in these studies spanning the range of 1.9--2.8~eV (roughly corresponding to wavelengths of 450--650~nm) \cite{Kikuchi_jjap_2004,Kishino_procspie_2007,Armitage_mrssp_2009,Chang_apl_2010, Sekiguchi_apl_2010,Armitage_nt_2010,Lin_apl_2010,Nguyen_nl_2011,Bavencove_nt_2011, Kunert_nt_2011,Albert_apl_2012,Kehagias_nt_2013}.
Note that there is only a minor difference of 10~meV in emission energy between our measurements at 10~K and at room temperature. The absence of a clear temperature dependence of the emission energy for the (In,Ga)N band as compared with the excitonic near-band edge (NBE) luminescence of GaN as well as the larger FWHM of the (In,Ga)N band at low temperature suggest the presence of compositional inhomogeneities leading to carrier localization. Calculations of the transition energy for the (In,Ga)N insertions in a one-dimensional approximation \cite{1DPoisson} using the thickness of 3~nm and In content of 26~\% derived from XRD are in good agreement with the observed emission energy. However, in light of our comparison between such calculations and the emission energies obtained for samples grown at different temperatures (and thus with varying In content) presented in \cite{Wolz_nt_2012}, this agreement must be considered to be incidental. The spectrum at 10~K shows an additional weak contribution around 2.8 eV that will be discussed later. 

The GaN luminescence originates from the NBE emission, but includes an additional contribution from the donor-acceptor pair (DAP) transition for the measurement at 10~K. The latter emission is due to a Mg memory effect in the used growth chamber resulting from experiments including this dopant which acts as an acceptor in GaN and is typical for samples from this MBE system.

Figure~\ref{fig:ingan_m81310ens}(b) shows room-temperature CL images of the (In,Ga)N band superimposed on the corresponding cross-sectional SEM image. They were recorded in spectral windows around the centre (2.2~eV) and the high-energy side (2.5~eV) of the (In,Ga)N band. These images confirm that the luminescence indeed comes from the area where the (In,Ga)N insertions are expected according to the design of the sample. Note that in CL carriers diffusing away from the position of the incident electron beam are still attributed to this position, which reduces the spatial resolution. At 2.2~eV, i.e., the lower of the two energies, the signal originates from a larger area. Carriers diffusing to the insertions tend to recombine at the lowest available energy states or potential minima. In contrast, a direct excitation of the insertions can lead to recombination also at higher energies, which is probably due to a filling of the lowest states under the increased excitation. As a consequence of the diffusion effect, the spatial resolution of the CL images is actually enhanced when detecting on the high energy side of the (In,Ga)N band.

\subsection{Single NW cathodoluminescence}

\begin{figure*}
\centering
\includegraphics*[width=15cm]{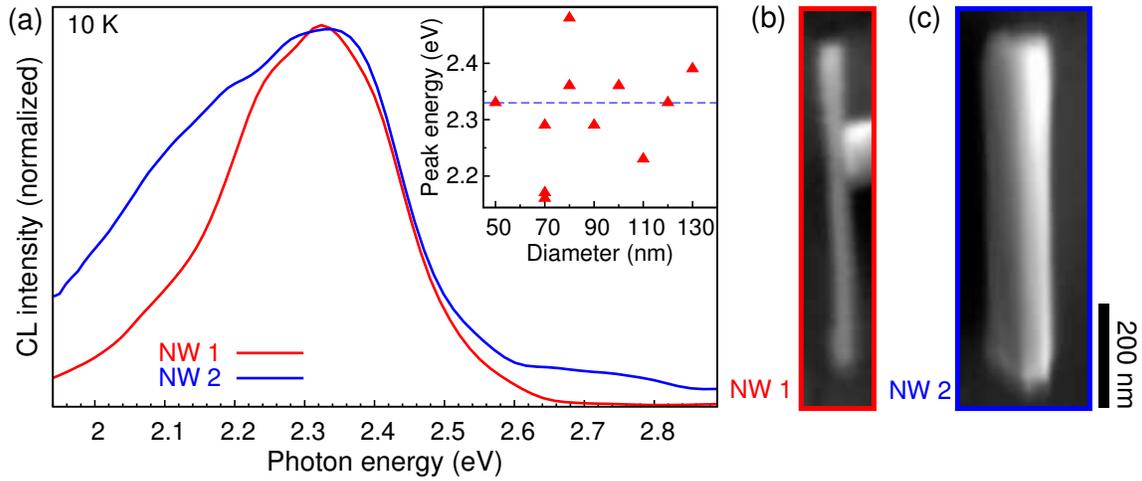} 
\caption{\label{fig:ingan_diameter}(a) Comparison of the CL spectra of two NWs with similar peak energies but different diameters. Diameters of 50 and 120~nm can be determined for the two NWs from their SEM images depicted in (b) and (c). The inset in (a) shows a plot of the emission energy versus diameter for 11 NWs; the dashed line marks the emission energy of NW~1 and NW~2.}
\end{figure*}

From previous studies, it is known that the large FWHM of the (In,Ga)N emission in NW ensembles partly results from a variation in the emission energy between individual NWs \cite{Lahnemann_prb_2011,Bavencove_nt_2011,Lin_apl_2010,Kehagias_nt_2013}. In \cite {Lahnemann_prb_2011}, it is shown that the peak energy of the (In,Ga)N band for single NWs can vary over a range of about 400~meV. Similar results are obtained for the present sample. The question arises, whether this spread of emission energies is correlated with the distribution of NW diameters. For GaN NWs, a diameter dependent influence of the sidewall diffusion of adatoms on the growth rate has been observed \cite{Debnath_apl_2007}. Therefore, assuming different diffusion rates of Ga and In atoms, an influence of the NW diameter on the dimensions and In content of the insertions could be expected. Also, the strain state should depend on the diameter. All these factors influence the emission energy of extended states in the QW. 

However, figure~\ref{fig:ingan_diameter}(a) shows that for our sample there is no correlation of the peak energy with the NW diameter in measurements on individual NWs at 10~K. The graph presents CL spectra obtained from two NWs both with the (In,Ga)N emission centred at 2.33~eV, but with diameters of 50 and 120~nm, respectively. The FWHM is 280~meV for the thinner NW and 450~meV for the thicker NW. Furthermore, the inset in figure~\ref{fig:ingan_diameter}(a) plots the peak energy of the (In,Ga)N emission versus the NW diameter for 11 investigated NWs and corroborates this conclusion.

The absence of a correlation between emission energy and NW diameter indicates that localization centres resulting from fluctuations of the In content within the insertions play a prominent role in determining their emission properties.

\subsection{Spatially and spectrally resolved emission within single NWs}

It is a common notion to talk of ``defect-free'' NWs. Indeed, dislocation densities are significantly reduced compared to planar heteroepitaxy \cite{Ristic_prb_2003,Yan_natphot_2009}, and also the emission related to native point defects might be less significant \cite{Brandt_prb_2010}. However, already the coalescence of neighboring NWs in dense arrays might introduce stacking faults and dislocations~\cite{Consonni_apl_2009,Grossklaus_jcg_2013}. Furthermore, reduced growth temperatures are required for the incorporation of In. This reduced growth temperature may increase the point defect density and can facilitate the nucleation of zinc-blende (ZB) segments \cite{Renard_apl_2010,Jacopin_jap_2011,Wolz_nt_2012,Kehagias_nt_2013}. The manifestation in the luminescence spectra of both intrinsic and extrinsic point defects as well as ZB segments will be presented in the following paragraphs. To this end, the luminescence distribution within single NWs is spatially resolved with the help of spectral line scans recorded along the axis of the NWs. Furthermore, the spatially resolved characteristics of the (In,Ga)N band is discussed.

Figure~\ref{fig:ingan_linescan} presents the results of a spectral line scan recorded at 10~K along an exemplary NW from the investigated sample. The individual spectra are plotted as a colour-coded map of position versus emission energy. Above the scan, four representative spectra are shown to highlight important features in the spectral map. The different contributions to this spectral map are discussed in the following.

\subsubsection{Contributions from point defects}

\begin{figure}
\centering
\includegraphics*[width=8.3cm]{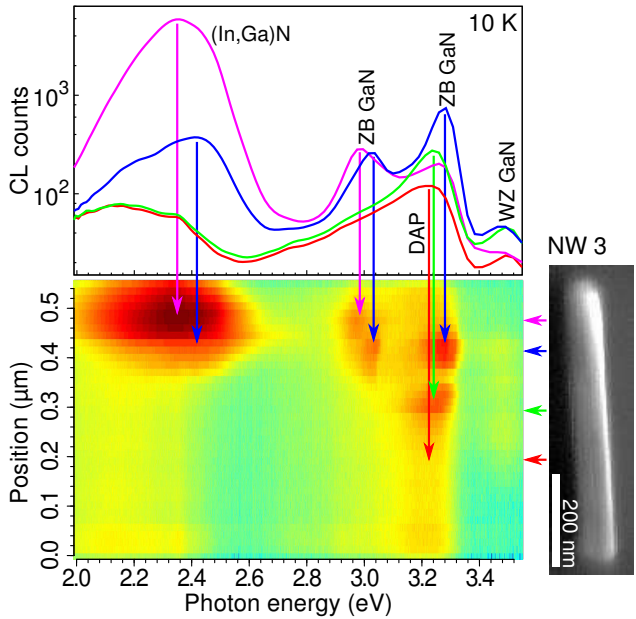}\hspace{4mm} 
\caption{\label{fig:ingan_linescan}CL spectral line scan on a single NW acquired at 10~K with the intensity plotted on a logarithmic colour scale and the corresponding SEM image set to the right. The spectra extracted from the line scan at the positions marked by coloured arrows and plotted above highlight contributions from point defects (red), as well as from ZB segments and a shift in the central emission energy along the axis of the stacked (In,Ga)N insertions.}
\end{figure}

A common feature of all CL spectral line scans recorded both at 300~K (not shown) and 10~K is an emission band between 2.0--2.4~eV that is visible all along the NW base (cf.\ figure~\ref{fig:ingan_linescan}). Its energy position is uncorrelated to that of the (In,Ga)N band. The extracted spectra in the top part of figure~\ref{fig:ingan_linescan} show that the emission intensity is about 2 orders of magnitude lower than for the (In,Ga)N band. The CL intensity remains almost constant until the (In,Ga)N band sets in. Such an emission at slightly lower energies than the main peak could stem from carriers diffusing to the (In,Ga)N insertions, but then a gradual increase of its intensity should be expected. Therefore, it is likely related to the yellow luminescence band, which is common for GaN layers grown by a variety of methods \cite{Reshchikov_jap_2005}. This emission band is generally attributed to point defects, the exact nature of which is under dispute \cite{Reshchikov_jap_2005}. A possible candidate is a transition between shallow donors and Ga vacancies acting as deep acceptors \cite{Neugebauer_apl_1996}, whereas a recent study attributes this emission to the (C$_\mathrm{N}$--O$_\mathrm{N}$)$^0$ deep donor complex \cite{Demchenko_prl_2013}. The presence of the yellow band is in contrast to the majority of other NW samples, for which an absence of this luminescence in PL measurements was observed and interpreted as an indication for a low density of native point defects \cite{Brandt_prb_2010,Geelhaar_stqe_2011}. While the NW base was grown at temperatures optimized for GaN, the appearance of the yellow luminescence in this specific sample might depend sensitively on factors such as the detailed growth conditions, defect interactions, or the purity of the material.

Additionally, the strong contribution from the DAP transition \cite{Reshchikov_jap_2005} in low-temperature spectra has already been mentioned. In figure~\ref{fig:ingan_linescan}, the DAP is seen to basically show a continuous distribution along the axis of the NW. However, in the upper part of the NW, it is superimposed by additional peaks as discussed in the following.

\subsubsection{Luminescence of structural defects}

The linescan in figure~\ref{fig:ingan_linescan} exhibits several emission peaks between 3.0 and 3.3~eV that originate from the upper part of the NW. As highlighted by the extracted spectra, these peaks are both offset in energy and more intense than the DAP transition. Many, but not all, of the measured NWs exhibit this spectral feature. We attribute these peaks to ZB segments acting as quantum wells in wurtzite GaN \cite{Jacopin_jap_2011,Lahnemann_prb_2012}, the presence of which has been repeatedly observed for NWs containing (In,Ga)N insertions \cite{Renard_apl_2010,Jacopin_jap_2011,Wolz_nt_2012,Kehagias_nt_2013}.
In line with the TEM observations in \cite{Kehagias_nt_2013}, the luminescence of ZB segments originates from a region just below the (In,Ga)N insertions or possibly from the first barriers between the lowermost insertions. The emission from such ZB quantum wells can be at lower energies than for bulk ZB GaN (3.27~eV), as it is governed by the quantum-confined Stark effect (QCSE). The observed emission range in figure~\ref{fig:ingan_linescan} corresponds to different thicknesses of the ZB segments between 1--2~nm (4--8 layers) \cite{Lahnemann_arxiv_2014}.

When looking at spectra integrated over one or more NWs, the emission from ZB segments can be masked by the DAP emission band at 3.27~eV and its phonon replicas [cf.\ figure~\ref{fig:ingan_m81310ens}(a)]. The CL line scan shows that under local excitation the emission from zinc-blende GaN segments clearly stands out against the DAP luminescence, a fact that showcases the power of this experimental method in disentangling spectrally overlapping contributions to the luminescence of nanostructures. This question could be addressed neither by spectra integrated over a larger area nor by monochromatic CL images, where one spatial or spectral dimension would be lacking.

\subsubsection{(In,Ga)N luminescence band}\label{sec:ingan-local}

In the following, we will focus on the (In,Ga)N band. From the spectra measured on the ensemble [Figure~\ref{fig:ingan_m81310ens}(a)], a rather large FWHM of more than 300~meV was determined. For single dispersed NWs, the emission energy of the (In,Ga)N band varies from NW to NW in the range of 2.1--2.5 eV (cf.\ inset in Fig~\ref{fig:ingan_diameter}(a) and \cite{Lahnemann_prb_2011}). This scatter in emission energies proves that the (In,Ga)N band of the NW ensemble is a superposition of these peaks and shows that it is dominant with respect to the yellow luminescence in the same spectral region that does not shift between NWs. However, the emission bands of individual NWs remain rather broad with an FWHM of 200--300~meV. In contrast, values for the FWHM on the order of 100--150~meV are reported from PL on planar (In,Ga)N QW structures emitting in the green spectral range  \cite{Lai_ape_2013,Li_ape_2013}. These features call for a closer investigation of the (In,Ga)N luminescence of single NWs in CL spectral maps.

\begin{figure}
\centering
\includegraphics*[width=8cm]{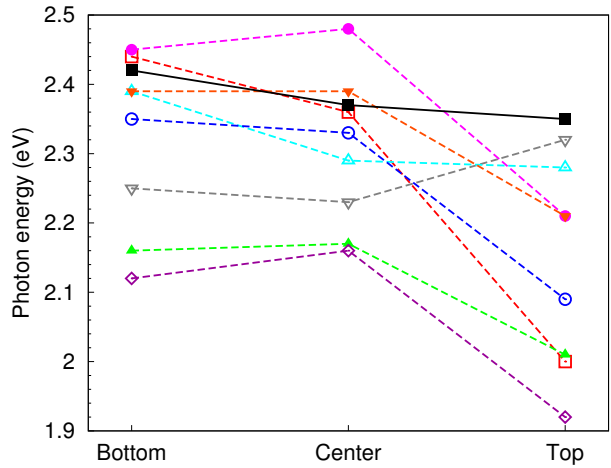}
\caption{\label{fig:ingan_energy-trend}Shift to lower emission energies along the stack of (In,Ga)N insertions. The development of the peak energy from the bottom to the top of the stack of (In,Ga)N insertions is plotted for three points from line scans on different NWs; the solid black line marks the data from figure~\ref{fig:ingan_linescan}.}
\end{figure}

\begin{figure*}
\centering
\includegraphics*[width=15cm]{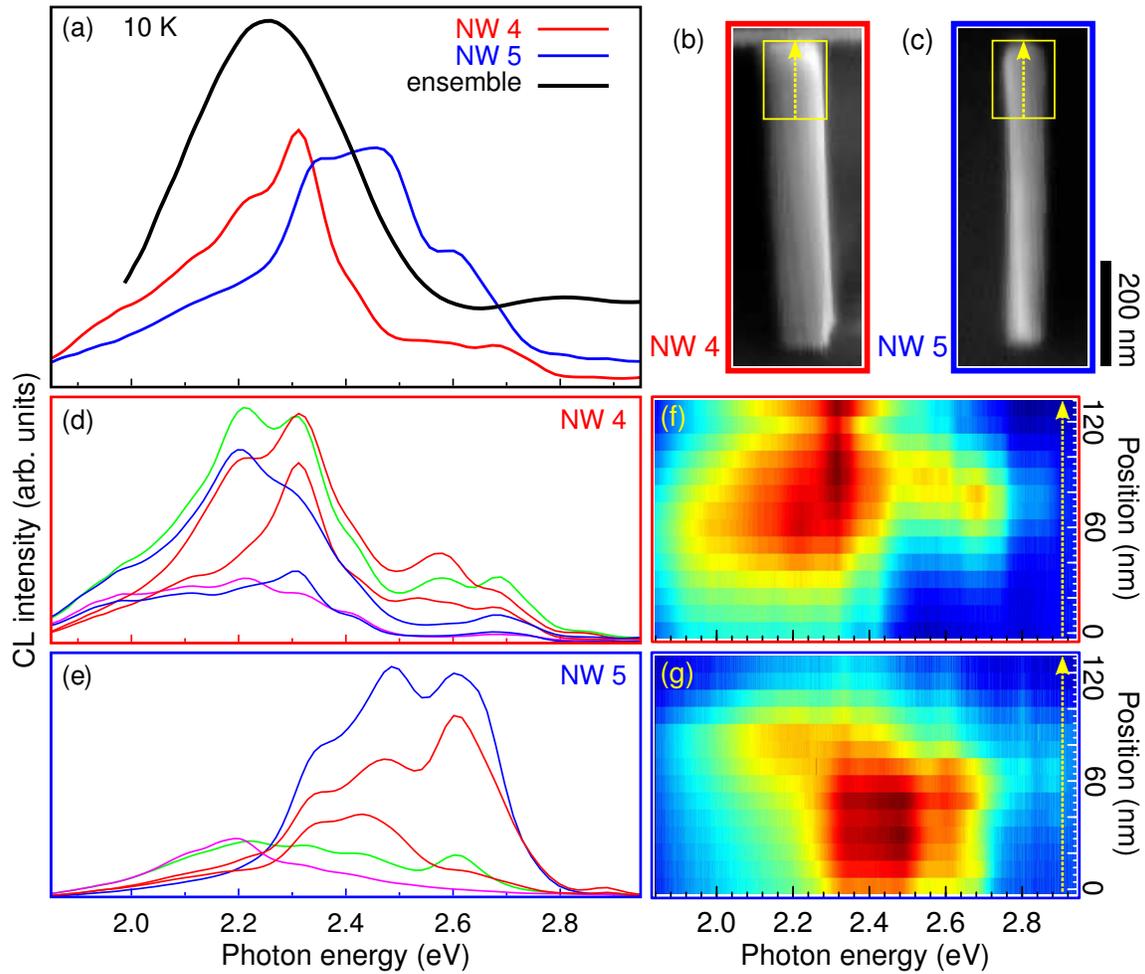} 
\caption{\label{fig:ingan_maps}(a) Comparison of the ensemble spectrum (thick black line) with spectra of two individual NWs (thinner lines). The spectra of NW~4 and NW~5 were mapped over the area marked in the SEM images in (b) and (c). Their spectra in (a) are integrated over this mapped area. Representative spectra from individual points of these maps are displayed in (d) for NW~4 and (e) for NW~5 to illustrate the inhomogeneity of the emission within a single NW; spectra displayed in the same colour are from points separated only in lateral direction, i.e., they highlight different contributions from the same (In,Ga)N insertion. To illustrate changes in the axial direction, the spectral maps are integrated in lateral direction and displayed as axial line scans in (f) for NW~4 and (g) for NW~5; note that these are plotted on a logarithmic colour scale, while the other spectra are plotted on a linear scale. The direction of the line scans is marked by arrows in (f) and (g) as well as in the SEM images in (b) and (c).}
\end{figure*}

The line scan in figure~\ref{fig:ingan_linescan} exhibits a slight shift of the (In,Ga)N emission energy towards lower energies along the stack of insertions. Two of the extracted spectra at the top of figure~\ref{fig:ingan_linescan} illustrate this redshift of the (In,Ga)N band. In fact, this trend is representative for most investigated NWs, as can be seen from the summary in figure~\ref{fig:ingan_energy-trend}. The emission energies at three points from the bottom, the centre and the top of the stack of (In,Ga)N insertions extracted from nine low-temperature CL line scans are depicted. For most investigated NWs, the shift is even more pronounced than in figure~\ref{fig:ingan_linescan} [see also figure~\ref{fig:ingan_maps}(g)]. The dimensions of these insertions are at the limit of the spatial resolution and thus individual insertions are not clearly resolved in SEM-based CL. Nevertheless, diffusing carriers should be trapped in the closest insertion \cite{Zagonel_nt_2012}, and therefore, the resolution is limited by the scattering volume of the electron beam with a diameter of about 30~nm at $V_\mathrm{acc}=5$~kV. Hence, the signal at a given point should include contributions from two to three insertions. The observed shift in emission energy, which can amount to 200~meV, is thus probably related to the uppermost insertions.

\begin{figure}
\centering
\includegraphics*[width=5.5cm]{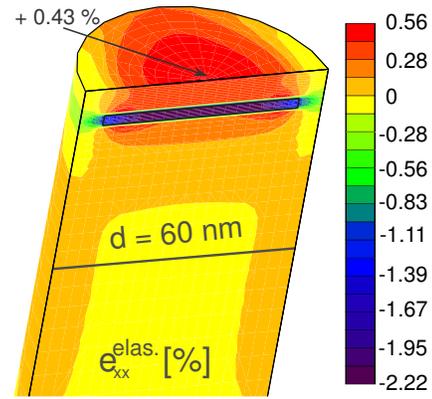}
\caption{\label{fig:ingan_fem}Finite-element simulation of the strain state at the growth front before starting the second insertion. A 3~nm thick insertion with an In content of 25~\% was embedded in the NW followed by a 7~nm thick barrier. The in-plane lattice constant at the centre of the wire is enlarged by 0.43~\% compared to the situation prior to growing the first insertion, where a completely unstrained GaN NW can be assumed.}
\end{figure}

A shift in emission energy along the NW axis was already observed by Tourbot \etal~\cite{Tourbot_nt_2012} on  a stack of three comparably thick (In,Ga)N insertions (15~nm thick insertions and 10~nm barriers), where the emission was resolved for individual insertions with the help of a TEM-based CL system. However, these authors observed the predominant part of the redshift already between the first and second insertion. Furthermore, the study showed that some insertions do not emit any light and that the lower part of the insertions shows a reduced luminescence intensity, which was attributed to the presence of non-radiative recombination centres. The differences in energy both within a single NW and between NWs were explained by the strain relaxation during growth resulting in the so-called compositional lattice pulling effect \cite{Tourbot_nt_2012}. To reduce the strain energy, a lower fraction of the available In atoms is incorporated initially followed by a gradual increase of the In content as the strain is partially relaxed \cite{Kong_nt_2012}. This effect could induce a gradient in the In content both in individual insertions and along the stack of insertions \cite{Tourbot_nt_2012}. In contrast, for rather thin insertions (4~nm) as in our sample, Kehagias \etal~\cite{Kehagias_nt_2013} did not observe the compositional pulling effect in their TEM analysis. To examine the potential magnitude of this effect in our structures, we calculated the strain state at the growth front prior to starting the second insertion using finite-element (FEM) simulations based on linear elasticity theory taking into account the full elastic anisotropy~\cite{Christiansen_pssa_1996}. As input, we took the mean structural parameters determined by XRD. The result is shown in figure~\ref{fig:ingan_fem}. Whereas the first insertion is grown on a fully relaxed GaN NW, the in-plane lattice-constant at the base for the second insertion is increased by 0.43~\% in the centre of the NW. This difference in lattice constants corresponds to roughly 4~\% more In in the alloy, although no direct conclusion can be drawn concerning the actual effect on the In incorporation. Nevertheless, it is thus plausible that a change in In content along the stack of insertions is indeed the explanation for the redshift in emission energy presented in figure~\ref{fig:ingan_energy-trend}.

To further extend the analysis of the (In,Ga)N emission, the results from a two-dimensional CL mapping on the top of two NWs are presented in figure~\ref{fig:ingan_maps}. First, in figure~\ref{fig:ingan_maps}(a), the integral emission spectra (sum of all spectra in the map) for NW~4 and NW~5 are compared to the ensemble spectrum. The mapped area is marked on the SEM images of the two NWs in figures~\ref{fig:ingan_maps}(b) and (c).

A complete graphical representation of the four-dimensional data set (x, y, energy, intensity) is impossible when the spectral maps contain multiple peaks at different positions. Therefore, a number of spectra from representative positions are shown in figures~\ref{fig:ingan_maps}(d) and (e) for NW~4 and NW~5, respectively. Figures~\ref{fig:ingan_maps}(f) and (g) depict the axial  line scans obtained by integrating all spectra from the maps in lateral direction (NW~4 corresponds to the one NW in figure~\ref{fig:ingan_energy-trend} that shows an opposite energy shift). The individual spectra from a single NW with six insertions show features that are common to several spectra as a result of the limited resolution. Still, distinct variations within a NW are found. The total number of different peak energies observed in the map of a single NW exceeds the number of embedded insertions, e.g., at least nine different peaks in the case of figure~\ref{fig:ingan_maps}(d). Among the individual spectra, those taken at the same height, i.e., separated only laterally, are plotted in the same colour. This is the case for two pairs of spectra in figure~\ref{fig:ingan_maps}(d) and for one pair in figure~\ref{fig:ingan_maps}(e). These pairs of spectra should originate from the same set of insertions. Nevertheless, quite different contributions to the spectra can be seen. These lateral fluctuations indicate the presence of localization centres induced by variations in the In content. A deconvolution of the spectra yields linewidths of 50--100~meV for the individual peaks, several of which can originate from the same insertion. For single (In,Ga)N insertions, Tourbot \etal~\cite{Tourbot_nt_2012} observed an FWHM of 200~meV in TEM-based CL measurements, which is comparable to our results. That is, the linewidth locally recorded on one insertion is of a similar magnitude as the integral luminescence of planar QW structures \cite{Lai_ape_2013,Li_ape_2013}.

Another observed feature is the emission at slightly higher energy than the main peaks visible in figures~\ref{fig:ingan_maps}(d) and (f) between 2.5 and 2.8~eV. A weak emission in this energy range is also visible in the low-temperature ensemble spectrum in figure~\ref{fig:ingan_m81310ens}. These peaks look quite similar to those of the ZB segments in figure~\ref{fig:ingan_linescan}, but they are redshifted by more than 200~meV and thus correspond to a ZB thickness of more than 3~nm \cite{Lahnemann_arxiv_2014}. 
The strong QCSE and thus further reduced wave-function overlap for such thick ZB segments renders this explanation unlikely. In line with our observations in~\cite{Lahnemann_prb_2011}, clusters with higher In content of nanometre dimensions are a more likely origin of these peaks. For sufficiently small clusters, the strong confinement-related blueshift can lead to such comparably high emission energies.

\subsection{Single NW photoluminescence}

\begin{figure}
\centering
\includegraphics*[width=8cm]{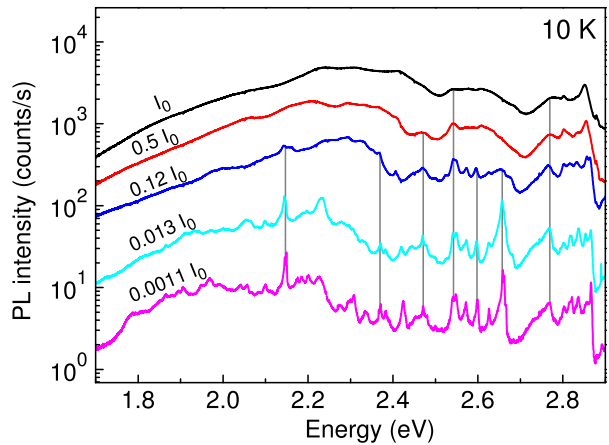} 
\caption{\label{fig:ingan_mupl} Excitation density dependent low-temperature $\mu$PL measurement on a single dispersed NW. The maximum excitation $I_0$ corresponds to 500~kW/cm$^2$. With decreasing excitation density, the continuous peak disintegrates into resolution limited peaks related to localized states distributed over a range of 1~eV. The vertical lines trace representative peaks to highlight that the emission energy is not shifted with increasing excitation density.}
\end{figure}

The significance of carrier localization for the emission properties of these (In,Ga)N insertions seen in the CL measurements is confirmed by $\mu$PL spectroscopy on single dispersed NWs with the insertions resonantly excited at about 3~eV. The low-temperature spectra of a representative NW are plotted in figure~\ref{fig:ingan_mupl} for different excitation densities spanning three orders of magnitude. For high excitation densities, the spectrum consists of a broad band centred around 2.3~eV and additional peaks in the range up to 2.9~meV. In CL we are always operating at relatively high excitation densities, and indeed, these features are similar to those in the CL spectra in figure~\ref{fig:ingan_maps}(d) discussed in the previous section. With reduced excitation density, it becomes evident that the whole emission band is built up of a superposition of sharp emission peaks resulting from carrier localization in the complex potential landscape of the insertions. The emission peaks are distributed over a range of 1~eV, which is even much wider than the linewidth of the emission band at higher excitation densities. As the insertions have a rather limited height of 3~nm and a diameter of 50--100~nm, the origin of the localization must lie in potential fluctuations on the scale of a few nanometres. In comparison, low-excitation PL measurements using nanoapertures on planar QWs show a significantly lower number of peaks over a spectral range of less than 100~meV~\cite{Schomig_prl_2004}. 

The majority of the emission peaks in figure~\ref{fig:ingan_mupl} can be tracked over several orders of excitation density without apparent spectral shift. This feature is a clear indication that we are dealing with the emission from excitons localized in single potential minima. In contrast, PL transients of similar samples evidence a DAP-like recombination of electrons and holes in separate potential minima \cite{Lahnemann_prb_2011,Cardin_nt_2013}. The latter states exhibit a much longer lifetime than localized excitons and dominate the time-resolved measurements, whereas the short-lived excitonic emission prevails in PL spectra under continuous excitation. In any case, the bottom-up fabrication of NW based QW structures seems to induce larger inhomogeneities in the alloy composition of the insertions than for planar QWs. Along this line, Kehagias \etal~\cite{Kehagias_nt_2013} have shown that for lower growth temperatures of the (In,Ga)N insertions, the inhomogeneity of the alloy is further increased and the width of the (In,Ga)N band for the NW ensemble can amount to 800~meV. A certain degree of localization might well be necessary to avoid the possibility of detrimental carrier recombination at the NW surface in line with what is discussed for dislocations in planar layers \cite{Oliver_jpd_2010}. However, strong inhomogeneities impede the desired tunability of emission properties \cite{Wolz_nt_2012}.

\section{Conclusions}

In the present study, we have investigated GaN NWs grown in the self-induced growth mode, which contain a stack of six thin (In,Ga)N insertions. CL images obtained on a cross-section of the NW ensemble confirm that the spatial origin of the main luminescence band coincides with the position of these insertions. Turning to CL spectral imaging, we could identify emission not only from the (In,Ga)N band, but also from point defects, namely the DAP transition and yellow luminescence. Furthermore, we observe luminescence related to ZB segments acting as QWs in the WZ GaN.

For the (In,Ga)N emission, we could observe different mechanisms contributing to the broad linewidth of the luminescence band. First, the emission energy varies between NWs. However, we find that the emission energy of the (In,Ga)N band does not correlate with the NW diameter.

Second, the emission energy varies among individual insertions in a single NW. Spectral line scans along the axis of a number of NWs show a redshift of the emission energy along the stack of (In,Ga)N insertions (from bottom to top). This result indicates that we may have a gradual change of the In content caused by the compositional pulling effect.

Third, carrier localization at potential fluctuations within single insertions induced by a variation of the In composition plays an important role. This localization appears to be more pronounced than in planar QWs and is a major factor affecting the emission of (In,Ga)N insertions embedded in GaN nanowires grown in the self-induced growth mode.

The different levels of inhomogeneity pose a challenge in attaining a sufficient control over the emission characteristics of these structures. For NWs etched from a layer in the top-down approach, CL measurements show a narrowing of the emission band compared to the original layer \cite{Bruckbauer_nt_2013}. Therefore, the pronounced inhomogeneity is a consequence of the bottom-up process. The ensemble broadening is particularly pronounced for NWs grown in the self-induced growth mode and might be reduced for more homogeneous NW ensembles grown by selective area growth. In the scope of the growth conditions for heterostructures produced in bottom-up mode investigated so far, the shift in emission energy within NWs as well as the inhomogeneity within single insertions seem to be more difficult to influence. Further research needs to address such questions.

Our results show that SEM-based CL can provide useful information even for these nanoscale structures that challenge the spatial resolution limits of the method. In comparison to TEM-based CL measurements, both the sample preparation and measurements are less sophisticated. The additional dimension in spectral images facilitates the distinction of emission from point defects, on the one hand, and quantum wells---be it intended (In,Ga)N insertions or unintentional ZB segments---on the other hand, thus highlighting the power of this experimental technique.

\ack

The authors would like to thank Sergio Fern\'a{}ndez-Garrido for a critical reading of the manuscript. This work has been partly funded by the German government BMBF project MONALISA (contract no. 01BL0810) as well as by the European Commission (FP7-NMP-2013-SMALL-7) under grant agreement no. 604416 (DEEPEN).

\bibliography{Laehnemann_InGaN-Martin_14}

\end{document}